\documentclass[preprint,showpacs,preprintnumbers,amsmath,amssymb]{revtex4}

\usepackage{graphicx}
\usepackage{dcolumn}
\usepackage{bm}
\usepackage[]{amssymb,epsfig,color}
\newcommand{\be}{\begin{equation}}
\newcommand{\ee}{\end{equation}}
\begin{document}
\title{Nonlinear dynamics of two coupled nano-electromechanical resonators}
\author{L. Chotorlishvili$^{1,3}$, A.Ugulava$^{2}$, G.Mchedlishvili$^{2}$, A. Komnik$^{3}$, S. Wimberger$^{3}$, J. Berakdar$^{1}$}
\affiliation{1 Institut f$\ddot{u}$r Physik, Martin-Luther
Universit$\ddot{a}$t, Halle-Wittenberg,
             Heinrich-Damerow-Str. 4, 06120 Halle, Germany\\
2 Physics Department of the Tbilisi State University, Chavchavadze av. 3, 0128, Tbilisi, Georgia \\
3 Institut f$\ddot{u}$r Theoretische Physik Universit$\ddot{a}$t
Heidelberg, Philosophenweg 19, 69120 Heidelberg, Germany}

\begin{abstract}
As a model of coupled nano-electromechanical resonantors we study
two nonlinear driven oscillators with  an arbitrary
coupling strength between them. Analytical expressions are derived for
the oscillation amplitudes as a function of the driving frequency
and for the energy transfer rate between the two oscillators. The
nonlinear restoring forces induce the expected nonlinear resonance
structures in the amplitude-frequency characteristics with asymmetric
resonance peaks. The corresponding multistable behavior is shown
to be an efficient tool to control the energy transfer arising
from the sensitive response to small changes in the driving
frequency. Our results imply that the nonlinear response can be
exploited to design precise sensors for mass or force detection
experiments based on nano-electromechanical resonators.
\end{abstract}

\maketitle
\section{Introduction}
\label{intro}

Currently experimental efforts are devoted to the fabrication of
nanoscale resonators with a  precise control of their behavior.
Such nanoscale resonators are ideal prototype systems for testing
fundamental physical concepts, such as entanglement and quantum
correlations \cite{Raimond}. By now several types of resonators
were successfully considered, like optical two level atoms in
quantum cavities \cite{Bruss}, artificial Josephson junction
qubits \cite{Shevchenko}, atoms seized in ion traps
\cite{iontrap}, or nano-optomechanical devises \cite{nanoopt}.
There is a new trend towards nano-electromechanical resonators.
They are widely studied from both the quantum-mechanical
\cite{iontrap,nanoel} and from a classical point of view
\cite{Lifshitz,Cross,Bromberg,Craighead,Buks,Rugar,Masmanidis,Karabalin}.
These devices are approximately  200 nm in size and consist of
three layers of gallium arsenide (GaAs): an $n$-doped layer of
width 100 nm is stacked within an insulating layer of 50 nm and a
$p$-doped layer of 50 nm \cite{Karabalin}. The resonators can be
controlled by electric fields via the piezoelectric effect, which
fix their mechanical strain \cite{Masmanidis}. Along with a single
resonator, one can consider coupled resonators driven by an
additional external field. Coupled resonators show different
dynamical regimes dependent on the interplay of their coupling
strength and the driving. For the case of moderate coupling
between two resonators this problem was already addressed in a
recent paper by Karabalin {\em et al.} \cite{Karabalin}. They
showed that the linear and weakly nonlinear response of one
oscillator can be modified by driving the second oscillator. A
complicated frequency-sweep response curve was obtained
numerically when both oscillators are driven into the strongly
nonlinear regime.

In this paper we study two nonlinear oscillators allowing for an
arbitrary coupling strength between them with a possibility of
driving both with the same frequency but different amplitudes. The
coupling strength between the oscillators is quantified in terms
of the connectivity parameter defined below in Sect. II. We derive
general analytical expressions for the amplitude-frequency
characteristics valid for arbitrary (weak as well as strong)
connectivity. We analyze the redistribution of energy between the
two resonators injected into the system via the external driving
fields. We quantify stable and unstable dynamical regimes, with
special focus on the nonlinear response of the system. Our
predictions point to possible new applications of nanoscale
resonators exploiting their sensitivity in response to external
fields and perturbations. In particular they may be used as
sensors for tiny forces or masses \cite{Roukes,Fischer} which lead
to a shift in their resonance frequencies to be identified in the
sensitive nonlinear response regime.

Our paper is organized as follows: in the next section, we discuss
our fundamental model of two coupled driven oscillators. In Sect.
III we study the mode frequency shifts and relaxation effects for
a non-resonant driving, while in Sect. IV we address the resonant
case with a special focus on the nonlinear shifts of the mode
frequencies. In the following sections we investigate the
frequency response function and the key problem of the energy
redistribution between the oscillators, before concluding in Sect.
VIII.

\section{Model}
We consider two nanomechanical oscillators described by the coordinates $x_{1,2}$ in the framework of the model outlined in \cite{Karabalin}. The corresponding dynamical equations can be written down in the following form:
$$\ddot{x}_{1}+\omega_{1}^{2}x_{1}+D(x_{1}-x_{2})=\varepsilon M,$$
\begin{equation}
\ddot{x}_{2}+\omega_{2}^{2}x_{2}+D(x_{2}-x_{1})=\varepsilon N.
\end{equation}
$$\varepsilon M=\varepsilon M(x_{1},\dot{x}_{1},t)=-2\gamma_{1}\dot{x}_{1}-\chi_{1}x_{1}^{3}+F_{1}\cos\Omega t,$$
$$~~~~~~~~~~~~~~~~~\varepsilon N=\varepsilon N(x_{2},\dot{x}_{2},t)=-2\gamma_{2}\dot{x}_{2}-\chi_{2}x_{2}^{3}+F_{2}\cos\Omega t,~~~~~~~~~~~~~~~~~~~~(1^{'})$$
$$\varepsilon\ll1.$$
Here $\omega_{1}$ and $\omega_{2}$  are frequencies of the
individual resonators, $\gamma_{1}$ and $\gamma_{2}$ are the
dissipation coefficients, $\chi_{1}$ and $\chi_{2}$ are
nonlinearity parameters, $F_{1}$ and $F_{2}$ are amplitudes of the
external harmonic forces applied to the resonators, $\Omega$ is
the frequency of these forces and $D$ is the coefficient of the
linear coupling between the resonators. As usual, we assume the
right hand side of Eqs.~(1) to be small perturbations, see also
\cite{Karabalin,Blaqueire,Hand, Steven,Landau}.

%
%

We summarize the canonical solution of the unperturbed coupled system first. Its dynamics follows the simple equations
$$\ddot{x}_{1}+\omega_{1}^{2}x_{1}+D(x_{1}-x_{2})=0,$$
\begin{equation}
\ddot{x}_{2}+\omega_{2}^{2}x_{2}+D(x_{2}-x_{1})=0.
\end{equation}
The transition from coupled oscillations to the mode
oscillations can be done via the following transformation
\cite{Landau},
$$x_{1}=q_{1}+q_{2},$$
\begin{equation}
x_{2}=-K^{-1}q_{1}+Kq_{2},
\end{equation}
where
\begin{equation}
K=-\frac{1}{\sigma}(1+\sqrt{1+\sigma^{2}}),~~K^{-1}=\frac{1}{\sigma}(1-\sqrt{1+\sigma^{2}}),~~KK^{-1}=1.
\end{equation}
\begin{equation}
\sigma=\frac{2D}{|\omega_{1}^{2}-\omega_{2}^{2}|}.
\end{equation}
We call the parameter $\sigma$ describing the coupling strength between the oscillators {\it connectivity}. From now on we assume that $\omega_{2}>\omega_{1}$.

The mode oscillations have the frequencies
\begin{equation}
\nu_{1,2}^{2}=\tilde{\omega}_{+}^{2}\mp\omega_{-}^{2}\sqrt{1+\sigma^{2}},
\end{equation}
where
$$\tilde{\omega}_{+}^{2}=\frac{\tilde{\omega}_{1}^{2}+\tilde{\omega}_{2}^{2}}{2},$$
$$\tilde{\omega}_{1,2}^{2}=\omega_{1,2}^{2}+D,$$
\begin{equation}
\omega_{-}^{2}=\frac{\omega_{2}^{2}-\omega_{1}^{2}}{2} \, .
\end{equation}
$\tilde{\omega}_{1,2}$ are  partial frequencies. We would like to stress that
the value of the connectivity $\sigma$ depends not only on the
linear coupling term $D$, but on the proximity of the free oscillation frequencies $\omega_{1}$ and $\omega_{2}$. In the limit of a weak connectivity $(\sigma\ll1)$ the frequencies $\nu_{1,2}$ tend to the partial frequencies
$\tilde{\omega}_{1,2}$, while in the limit of strong connectivity
$(\sigma\gg1)$,
\begin{equation}
\nu_{1}^{2}\simeq\omega_{+}^{2},~~~\nu_{2}^{2}\simeq\tilde{\tilde{\omega}}_{+}^{2},
\end{equation}
where
$$\omega_{+}^{2}=\frac{\omega_{1}^{2}+\omega_{2}^{2}}{2},~~~\tilde{\tilde{\omega}}_{+}^{2}=\frac{\omega_{1}^{2}+\omega_{2}^{2}}{2}+2D.$$
Obviously, in the limit $\sigma\gg 1$ the mode
frequency separation attains the maximal possible value which is  equal
to $2D$.

In the case of  a finite driving $F_{1,2}\neq 0$ but in the linear ($\chi_{1,2}=0$) dissipationless regime ($\gamma_{1,2}=0$) the particular solutions of the dynamical equations (1) are given by
%
%
$$x_{1}=A_{1}\cos\Omega t,~~x_{2}=A_{2}\cos\Omega t,$$
\begin{equation}
A_{1,2}=\frac{F_{1,2}(\tilde{\omega}_{2,1}^{2}-\Omega^{2})+F_{2,1}D}{d^{2}},
\end{equation}
here
$$\frac{1}{d^{2}}=\frac{1}{(\Omega^{2}-\nu_{1}^{2})(\Omega^{2}-\nu_{2}^{2})}=$$
\begin{eqnarray}            \label{eq11}
=\frac{1}{\Omega^{2}(\nu_{2}^{2}-\nu_{1}^{2})}\Bigl(\frac{\nu_{1}^{2}}{\nu_{1}^{2}-\Omega^{2}}-\frac{\nu_{2}^{2}}{\nu_{2}^{2}-\Omega^{2}}\Bigr)
\end{eqnarray}
and $A_{1,2}$ are the amplitudes of the induced resonator oscillations. They
increase resonantly when the frequency of the external driving tends closer to either of $\nu_{1,2}$.

With the above solutions at hand the influence of the damping term as well as of the nonlinear
corrections can be taken into account with the help of a standard substitution in the resonant denominator \cite{Landau}:
$$\frac{1}{\nu_{1,2}^{2}-\Omega^{2}}\rightarrow\frac{1}{2\nu_{1,2}(\nu_{1,2}-\Omega)}\rightarrow\frac{1}{2\nu_{1,2}(\nu_{1,2}+\delta _{1,2}+i\Gamma_{1,2}-\Omega)}\rightarrow$$
\begin{eqnarray}                \label{subst}
\rightarrow\frac{1}{2\nu_{1,2}\sqrt{(\nu_{1,2}+\delta_{1,2}-\Omega)^{2}+\Gamma_{1,2}^{2}}},
\end{eqnarray}
where $\Gamma_{1,2}$ are the mode relaxation rates, $\delta_{1,2}$
are the nonlinear corrections to the  mode frequencies that depend
on the oscillation amplitudes $A_{1,2}$. We would like to point out that the
substitutions of Eq.~(\ref{subst}) are correct only if the nonlinearity is not too
strong and the decay rate is not too high $(\nu_{1,2}\gg\delta
_{1,2};~~\Gamma_{1,2})$. Explicit expressions for $\delta _{1,2}$
and $\Gamma_{1,2}$, in terms of the system parameters will be
given in the next section.

\section{Nonlinear shift of the mode frequencies and consequences
of the relaxation terms. The non-resonant case}

We turn back to the perturbed system of Eq.~(1) making the
preliminary assumption that the resonance condition does not hold at
the mode frequencies $\nu_{1,2}\neq\Omega$. It is clear that in
this particular case, the role of the external force is
negligible. We concentrate at first on the influence of the
relaxation and of the nonlinearity on the oscillation of the coupled resonators.

To study the equations (1), we use a modified method of a slowly
varying amplitudes, \cite{Blaqueire}. Taking into consideration
the transformation (3), we reduce the unperturbed system to the
mode oscillations and write down the solution in the form:
$$x_{1}(t)=A_{1}(t)\sin[\nu_{1}t+\alpha_{1}(t)]+A_{2}(t)\sin[\nu_{2}t+
\alpha_{2}(t)],$$
\begin{eqnarray}                  \label{xx}
x_{2}(t)=-K^{-1}A_{1}(t)\sin[\nu_{1}t+\alpha_{1}(t)]
+KA_{2}(t)\sin[\nu_{2}t+\alpha_{2}(t)],
\end{eqnarray}
where $A_{1,2}(t)$, $\alpha_{1,2}(t)$ are slowly varying amplitudes and phases, respectively. The variables $\dot{A}_{1,2}(t)$ are the first-order
infinitesimal variables and therefore the terms proportional to the second-order derivatives $\ddot{A}_{1,2}(t)$ can be omitted upon an insertion of the above relations into  Eq.~(1). Following the standard procedures, after straightforward but laborious calculations, we find  for
the slowly varying amplitudes and the phases
$$\frac{dA_{1}}{dt}=\frac{1}{4\nu_{1}}\frac{\sigma}{\sqrt{1+\sigma^{2}}}(-K^{-1}P_{1}+Q_{1}),$$
$$\frac{dA_{2}}{dt}=-\frac{1}{4\nu_{2}}\frac{\sigma}{\sqrt{1+\sigma^{2}}}(KP_{2}+Q_{2}),$$
$$A_{1}\frac{d\alpha_{1}}{dt}=-\frac{1}{4\nu_{1}}\frac{\sigma}{\sqrt{1+\sigma^{2}}}(-K^{-1}P_{3}+Q_{3}),$$
\begin{equation}
A_{2}\frac{d\alpha_{2}}{dt}=\frac{1}{4\nu_{2}}\frac{\sigma}{\sqrt{1+\sigma^{2}}}(KP_{4}+Q_{4}),
\end{equation}
where
$$P_{1}=\frac{1}{4\pi^{3}}\int_{0}^{2\pi}\int_{0}^{2\pi}\int_{0}^{2\pi}M\cos\xi d\xi d\eta
d\zeta,
~~Q_{1}=\frac{1}{4\pi^{3}}\int_{0}^{2\pi}\int_{0}^{2\pi}\int_{0}^{2\pi}N\cos\xi
d\xi d\eta d\zeta,$$
$$P_{2}=\frac{1}{4\pi^{3}}\int_{0}^{2\pi}\int_{0}^{2\pi}\int_{0}^{2\pi}M\cos\eta d\xi d\eta
d\zeta,
~~Q_{2}=\frac{1}{4\pi^{3}}\int_{0}^{2\pi}\int_{0}^{2\pi}\int_{0}^{2\pi}N\cos\eta
d\xi d\eta d\zeta,$$
$$P_{3}=\frac{1}{4\pi^{3}}\int_{0}^{2\pi}\int_{0}^{2\pi}\int_{0}^{2\pi}M\sin\xi d\xi d\eta
d\zeta,
~~Q_{3}=\frac{1}{4\pi^{3}}\int_{0}^{2\pi}\int_{0}^{2\pi}\int_{0}^{2\pi}N\sin\xi
d\xi d\eta d\zeta,$$
\begin{equation}
P_{4}=\frac{1}{4\pi^{3}}\int_{0}^{2\pi}\int_{0}^{2\pi}\int_{0}^{2\pi}M\sin\eta
d\xi d\eta d\zeta,
~~Q_{4}=\frac{1}{4\pi^{3}}\int_{0}^{2\pi}\int_{0}^{2\pi}\int_{0}^{2\pi}N\sin\eta
d\xi d\eta d\zeta.
\end{equation}
For $M$ and $N$, determined by Eq.($1^{\prime}$), after inserting
$x_{1}(t)$ and $x_{2}(t)$ from (\ref{xx}), we obtain
$$M(\xi,\eta,\zeta)=-\gamma_{1}(A_{1}\nu_{1}\cos\xi+A_{2}\nu_{2}\cos\eta+
A_{1}\Omega\sin\zeta)-\alpha_{1}(A_{1}\sin\xi+A_{2}\sin\eta+A_{1}\cos\zeta)^{3},$$
$$N(\xi,\eta,\zeta)=-\gamma_{2}(-A_{1}K^{-1}\nu_{1}\cos\xi+A_{2}K\nu_{2}\cos\eta-
A_{2}\Omega\sin\zeta)-$$
$$-\alpha_{2}(-A_{1}K^{-1}\sin\xi+A_{2}K\sin\eta+A_{2}\cos\zeta)^{3},$$
\begin{equation}
\xi=\nu_{1}t+\alpha_{1};~~~\eta=\nu_{2}t
+\alpha_{2};~~~\zeta=\Omega t.
\end{equation}
Inserting (15) into (14), after simple integration, from (13) one
infers
$$\frac{dA_{1}}{dt}=-\frac{1}{2}\Gamma_{1};~~~\frac{dA_{2}}{dt}=-\frac{1}{2}\Gamma_{2};$$
\begin{equation}
\frac{d\alpha_{1}}{dt}=\delta _{1};
~~~\frac{d\alpha_{2}}{dt}=\delta _{2};
\end{equation}
where
$$\Gamma_{1,2}=\frac{1}{2}\Bigl[\gamma_{1}\Bigl(1\pm\frac{1}{\sqrt{1+\sigma^{2}}}\Bigr)+\gamma_{2}\Bigl(1\mp\frac{1}{\sqrt{1+\sigma^{2}}}\Bigr)\Bigr],$$
\begin{equation}
\delta_{1,2}=\frac{3}{8}\cdot\frac{1}{\nu_{1,2}}\Bigl[\chi_{1}A_{1}^{2}
\Bigl(1\pm\frac{1}{\sqrt{1+\sigma^{2}}}\Bigr)+\chi_{2}A_{2}^{2}\Bigl(1\mp\frac{1}{\sqrt{1+\sigma^{2}}}\Bigr)\Bigr]
\end{equation}
are the  relaxation rates and the nonlinear shifts of the mode
frequencies. An interesting fact is that the shift of the mode frequencies
$\delta_{1,2}$ depends on the  square of the amplitudes
$A_{1,2}^{2}$ as a   consequence of the nonlinearity.

In the case of a  weak connectivity ($\sigma\ll 1$), from Eq.~(17)  we
deduce
$$\Gamma_{1,2}\simeq\gamma_{1,2},~~~\delta_{1,2}=\frac{3}{4}\frac{1}{\tilde{\omega}_{1,2}}\chi_{1,2}A_{1,2}^{2},$$
while in case of strong connectivity ($\sigma\gg 1$)
$$\Gamma_{1}=\Gamma_{2}\approx\frac{1}{2}(\gamma_{1}+\gamma_{2}),$$
\begin{equation}
\delta_{1}\approx\frac{3}{8}\sqrt{\frac{2}{\omega_{2}^{2}+\omega_{1}^{2}}}(\chi_{1}A_{1}^{2}+\chi_{2}A_{2}^{2}),
\end{equation}
$$\delta_{2}\approx\frac{3}{8}\sqrt{\frac{2}{\omega_{2}^{2}+\omega_{1}^{2}+4D}}(\chi_{1}A_{1}^{2}+\chi_{2}A_{2}^{2}).$$

Therefore, when $\sigma\gg 1$ the modes are damped with the equal rates. However, the nonlinear shifts of the mode frequencies are different ($\delta_{1}>\delta_{2}$).

\section{Frequency response function of two coupled resonators}

The amplitudes of the driven oscillations of the two coupled resonators are presented by the expressions (9) and (10). Reexpressed in terms of the connectivity $\sigma$, we can rewrite the amplitudes as
$$A_{1,2}= \frac{ F_{1,2}(\tilde{\omega}_{2,1}^{2}-\Omega^{2})+F_{2,1}\omega_{-}^{2}\sigma}{2\Omega^{2}}\times$$
\begin{equation}
\times\Bigl[\frac{\omega_{+}^{2}}{\omega_{-}^{2}}\frac{1}{\sqrt{1+\sigma^{2}}}\Bigl(\frac{1}{\nu_{1}^{2}-\Omega^{2}}-
\frac{1}{\nu_{2}^{2}-\Omega^{2}}\Bigr)-\Bigl(\frac{1}{\nu_{1}^{2}-\Omega^{2}}+\frac{1}{\nu_{2}^{2}-\Omega^{2}}\Bigr)\Bigr].
\end{equation}

In the limit of a weak connectivity
($\sigma\ll 1$) $A_{1,2}=2F_{1,2}/(\omega_{1,2}^{2}-\Omega^{2})$
whereas in the limit of a strong connectivity ($\sigma\gg 1$) we obtain
\begin{equation}
A_{1,2}=\frac{DF_{2,1}}{\Omega^{2}}\Bigl(\frac{1}{\Omega^{2}-\omega_{+}^{2}}+\frac{1}{\Omega^{2}-\tilde{\tilde{\omega}}_{+}^{2}}\Bigr).
\end{equation}

Nanomechanical resonators for intermediate values of the connectivity
$(\sigma\approx2)$ were studied numerically and experimentally in
\cite{Karabalin}. While the general analytical expression (19) is
derived for   arbitrary values of the connectivity and
therefore in the special case of a moderate connectivity the
solution recovers the amplitude-frequency characteristics obtained
in \cite{Karabalin}. Using the transformation (11), one can easily
modify (19) in order to add corrections describing the damping and
nonlinear terms. However, the expressions obtained in this way are
 rather involved (see Appendix A). That is why here we only
present the asymptotic expressions corresponding to the strong
connectivity limit ($\sigma\gg 1$):
\begin{equation}
A_{1,2}\simeq\frac{DF_{1,2}}{\Omega^{2}}\Bigl(\frac{1}{\omega_{+}\sqrt{(\omega_{+}+\delta_{1}-\Omega)^{2}+\Gamma_{1}^{2}}}
+\frac{1}{\tilde{\tilde{\omega}}_{+}\sqrt{(\tilde{\tilde{\omega}}_{+}+\delta_{2}-\Omega)^{2}+\Gamma_{2}^{2}}}\Bigr),
\end{equation}
where $\delta_{1,2}$ and $\Gamma_{1,2}$ are determined in (18).

As follows from (21), in the case of a strong connectivity the force
$F_{2}$ acting on the second oscillator, ``drives" the first one,
and vice versa -- $F_{1}$, acting on the first oscillator, ``drives"
the second one.

\begin{figure}[t]
 \centering
  \includegraphics[width=10cm]{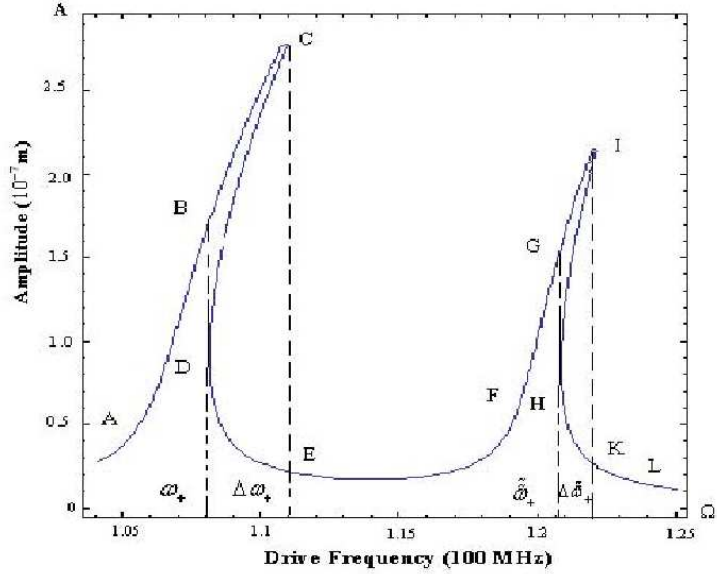}
  \caption{\label{fig1} Amplitude-frequency characteristics for the system of  two
strongly coupled oscillators, plotted using Eq.(21) and following
values of the parameters:
$F_{1}=F_{2},~~A=A_{1,2},~~\omega_{+}=1.07\cdot10^{8}~Hz,~~\tilde{\tilde{\omega}}_{+}=1.2\cdot10^{8}~Hz,~~
\delta_{1}=50.4\cdot10^{18}A^{2}~Hz,~~\delta_{2}=45.0\cdot10^{18}A^{2}~Hz,~~\Gamma_{1}=\Gamma_{2}=2.0\cdot10^{5}~Hz,
~~DF_{1,2}=81.6\cdot10^{21}mHz^{4}$.}
\end{figure}

The amplitude--frequency characteristic consists of  two tilted peaks with different heights, see Fig.~\ref{fig1}. The first peak corresponds to the
frequency $\omega_{+}$ and is definitely more pronounced than the second peak, corresponding to the frequency
$\tilde{\tilde{\omega_{+}}}~~(\tilde{\tilde{\omega_{+}}}\gg\omega_{+})$.
Furthermore, the first peak is more tilted due to the relation
$\delta_{1}>\delta_{2}$. The parts of the plot $CD$ and $IH$,
corresponding to unstable oscillations of the system, are dotted.
During upward/downward frequency sweeps of $\Omega$ one observes hysteretic behaviour around $\Delta \omega_{+}$ and
$\Delta\tilde{\tilde{\omega}}_{+}$ along the loops -$BCED$ and $GIKH$, respectively.\cite{Blaqueire,Landau} In
\cite{Karabalin} similar hysteretic loops in amplitude--frequency characteristics of coupled nonlinear oscillators were obtained
numerically and were confirmed experimentally for intermediate
values of the connectivity $\sigma\approx2$. In the unstable region
the system is extremely sensitive to the perturbations. This fact can
be used for the switching of the oscillation amplitude. After
reaching the  point $C$, the amplitude of the oscillation decreases sharply to the value $E$. Therefore, a simple and efficient switching protocol can be realized by tuning of the external field frequency only.

We would like to point out, that the domain of the amplitude frequency characteristics that should be utilized for switching belongs to the unstable area (see the frequency intervals
$B-C$ and $G-I$ on Fig.~1). Therefore, the system can jump to the lower state before reaching the summit of the unstable domain (point $C$).  If this happens the jump of the oscillation amplitude is smaller making difficult the experimental observation of the two different transport regimes.To circumvent  this problem  the frequency of the driving field should be changed adiabatically. Following
our approach here we seek the criteria of adiabaticity
that may be useful for different realizations of the system. A similar problem arises for example when studying
nonlinear resonant transport in cold atoms \cite{Richter}. The method of the slow varying amplitudes implies that the amplitude change rate should be slower than  the mode frequencies $\nu_{1,2}$.
Therefore, the rate of the oscillation amplitude change caused by
tuning  the frequency of the driving field is limited by the
following condition:
         \be \frac{dA_{1,2}(\Omega(t))}{dt}=\frac{dA_{1,2}(\Omega(t))}{d\Omega(t)}\frac{d\Omega(t)}{dt}<\nu_{1,2}
         A_{1,2}(\Omega(t)).\ee
The adiabaticity condition can then be simplified taking into account the explicit expressions for the amplitudes given in Eq.~(21). It is not difficult to show that in the vicinity of unstable areas
\be\frac{dA_{1,2}(\Omega(t))}{d\Omega(t)}\approx\frac{2A_{1,2}(\Omega(t))}{\Omega(t)}.\ee
Thus, for the adiabaticity criteria we finally obtain the
following estimation \be\frac{d\Omega(t)}{dt}<\mbox{min}
\big(\nu_{1,2}\big)\frac{\Omega(t)}{2}.\ee

\section{Nonlinear shift of the mode frequencies and the influence of
the relaxation terms. The resonant case}

Let us suppose that the harmonic force $F_{1}\cos\Omega t$ is tuned in resonance with one of the modes and that $F_{2}=0$. For this problem we derive equations for the slowly
varying amplitudes in a more straightforward way. Taking into
consideration the resonance condition and the transformation
(3), we can write down the solution of the equation set (1) in the following form:
$$x_{1}(t)=A_{1}(t)\sin\nu_{1}t+A_{2}(t)\cos\nu_{1}t+B(t)\sin(\nu_{2}t+\psi(t)),$$
\begin{equation}
x_{2}(t)=-K^{-1}(A_{1}(t)\sin\nu_{1}t+A_{2}(t)\cos\nu_{1}t+KB(t)\sin(\nu_{2}t+\psi(t)).
\end{equation}
After application of the standard method outlined in the last Section, for equations of slowly varying
amplitudes and phases we obtain
$$\frac{dA_{1}}{dt}=\frac{1}{4\nu_{1}}\frac{\sigma}{\sqrt{1+\sigma^{2}}}(-K^{-1}P_{1}^{(r)}+Q_{1}^{(r)}),$$
$$\frac{dA_{2}}{dt}=-\frac{1}{4\nu_{1}}\frac{\sigma}{\sqrt{1+\sigma^{2}}}(-KP_{2}^{(r)}+Q_{2}^{(r)}),$$
$$\frac{dB}{dt}=-\frac{1}{4\nu_{2}}\frac{\sigma}{\sqrt{1+\sigma^{2}}}(K^{-1}P_{3}^{(r)}+Q_{3}^{(r)}),$$
\begin{equation}
\frac{1}{B}\frac{d\psi}{dt}=\frac{1}{4\nu_{2}}\frac{\sigma}{\sqrt{1+\sigma^{2}}}(KP_{4}^{(r)}+Q_{4}^{(r)}),
\end{equation}
where
$$P_{1}^{(r)}=\frac{1}{2\pi^{2}}\int_{0}^{2\pi}\int_{0}^{2\pi}M\cos\eta d\xi
d\eta,
~~~~Q_{1}^{(r)}=\frac{1}{2\pi^{2}}\int_{0}^{2\pi}\int_{0}^{2\pi}N\cos\eta
d\xi d\eta,$$
$$P_{2}^{(r)}=\frac{1}{2\pi^{2}}\int_{0}^{2\pi}\int_{0}^{2\pi}M\sin\eta d\xi
d\eta,
~~~~Q_{2}^{(r)}=\frac{1}{2\pi^{2}}\int_{0}^{2\pi}\int_{0}^{2\pi}N\sin\eta
d\xi d\eta,$$
$$P_{3}^{(r)}=\frac{1}{2\pi^{2}}\int_{0}^{2\pi}\int_{0}^{2\pi}M\cos\xi d\xi
d\eta,
~~~~Q_{3}^{(r)}=\frac{1}{2\pi^{2}}\int_{0}^{2\pi}\int_{0}^{2\pi}N\cos\xi
d\xi d\eta,$$
\begin{equation}
P_{4}^{(r)}=\frac{1}{2\pi^{2}}\int_{0}^{2\pi}\int_{0}^{2\pi}M\sin\xi
d\xi d\eta,
~~~~Q_{4}^{(r)}=\frac{1}{2\pi^{2}}\int_{0}^{2\pi}\int_{0}^{2\pi}N\sin\xi
d\xi d\eta.
\end{equation}
$$\xi=\nu_{2}t+\psi,~~~~~~\eta=\Omega t,$$
$$M^{(r)}=M^{(r)}(x_{1},\dot{x}_{1},t)=-\gamma_{1}[\nu_{1}(A_{1}\cos\eta-\sin\eta)+B\cos\xi]-$$
$$-\chi_{1}(A_{1}\sin\eta+A_{2}\cos\eta+B\cos\xi)^{3}+F\cos\Omega t,$$
$$N^{(r)}=N^{(r)}(x_{2},\dot{x}_{2},t)=-\gamma_{2}[-K^{-1}\nu_{1}(A_{1}\cos\eta-A_{2}\sin\eta)+\nu_{2}B\cos\xi]-$$
\begin{equation}
-\chi_{2}[-K^{-1}(A_{1}\sin\eta+A_{2}\cos\eta)+KB\sin\xi]^{3}.
\end{equation}
Upon insertion of Eqs.(28) into (27) and after an integration, one
gets equations for the slowly varying amplitudes $A_{1},A_{2},B$
and $\psi$. The explicit form of these equations is given in
Appendix B. In the limit of a strong connectivity  $(\sigma\gg
1)$, the expressions simplify to
$$\frac{dA_{1}}{dt}=\frac{F}{4\Omega}-\gamma A_{1}-\frac{3\chi}{8\Omega}(A_{1}^{2}+A_{2}^{2}+2B^{2})A_{2},$$
$$\frac{dA_{2}}{dt}=-\gamma A_{2}+\frac{3\chi}{8\Omega}(A_{1}^{2}+A_{2}^{2}+2B^{2})A_{1},$$
\begin{equation}
\frac{dB}{dt}=-\gamma B.
\end{equation}
As is evident, a resonant external force $F_{1}\cos\Omega
t,~~(\Omega=\nu_{1}),$ for $\gamma=\chi=0$ leads to the simplest
form of instability (the secular instability), namely to the
linear growth of the oscillation amplitude $A_{1}=(F/4\Omega)t$.

We would like to point out that in the first three equations for the
variables $A_{1},~A_{2}$ and $B$, the right hand side  of the set of Eqs.~(29)
does not depend on the fourth variable $\psi$. Therefore, the set of
Eqs.~(29) can be solved self-consistently for the first three
variables.

In order to find the stationary values of the slowly varying
amplitudes and in order to examine the stability of these values, we utilize the following transformation:
\begin{equation}
A_{1}=\rho\cos\theta,~~~~~~A_{2}=-\rho\sin\theta \, .
\end{equation}
In the more convenient polar coordinates $\rho$ and $\theta$ we obtain
$$\frac{d\rho}{dt}=-\gamma\rho+\frac{F}{4\Omega}\cos\theta,$$
$$\frac{d\theta}{dt}=-\omega_{NL}-\frac{F}{4\Omega}\frac{\sin\theta}{\rho},$$
\begin{equation}
\frac{dB}{dt}=-\gamma B \, ,
\end{equation}
where $\omega_{NL}=\frac{3\chi}{8\Omega}(\rho^{2}+2B)$. By setting
the rhs of Eqs.~(31) equal to zero, we obtain equations for the
stationary values of amplitudes:
\begin{equation}
B_{0}=0,~~~s\cos\theta_{0}=\rho_{0},~~~s\sin\theta_{0}=-r\rho_{0}^{3} \, ,
\end{equation}
where
$s=\frac{F}{4\gamma\Omega},~~~r=\frac{3}{4}\frac{\chi}{\gamma\Omega}$.
To determine $\rho_{0}$, we eliminate the variable $\theta_{0}$
from the set of Eqs.~(31) and obtain a cubic equation with respect
to $x=\rho_{0}^{2}$:
\begin{equation}
x^{3}+\frac{x}{r^{2}}-\frac{s^{2}}{r^{2}}=0.
\end{equation}
Eq.~(33) is a reduced cubic equation. The number of real roots of
this equation depends on the sign of the discriminant:
\begin{equation}
D=\Bigl(\frac{1}{3r^{2}}\Bigr)^{3}+\Bigl(\frac{s^{2}}{2r^{2}}\Bigr)^{2}>0 \, ,
\end{equation}
which is positive in our case. That is why Eq.~(33) has a real
root. Real roots of Eq.~(33) can be identified easily with the
help of well-known Cardano's formula \cite{Korn}. However, as it
will become evident below, they are not necessary  for a further
specification of the expressions for the stationary points need
for  the study of the stability conditions.

To address the question concerning the  stability of  the stationary points more
precisely we linearize the set of Eqs.~(29) in the vicinity of the
stationary points
$A_{1}^{(0)}=\rho_{0}\cos\theta_{0},~~A_{2}^{(0)}=\rho_{0}\sin\theta_{0}$
and $B_{0}=0$, and obtain:
$$\delta \dot{A}_{1}=-\gamma(1+2rA_{1}^{(0)}A_{2}^{(0)})\delta
A_{1}-\gamma r(A_{1}^{(0)2}+3A_{2}^{(0)2})\delta A_{2},$$
\begin{equation}
\delta \dot{A}_{2}=\gamma r(A_{2}^{(0)2}+3A_{1}^{(0)2})\delta
A_{1}-\gamma (1-2rA_{1}^{(0)}A_{2}^{(0)})\delta A_{2}.
\end{equation}
Alternatively, by taking into consideration the transformation
(30):
$$\delta \dot{A}_{1}=R_{11}\delta A_{1}+R_{12}\delta A_{2},$$
\begin{equation}
\delta \dot{A}_{2}=R_{21}\delta A_{1}+R_{22}\delta A_{2},
\end{equation}
where
\begin{equation}
R=\left(\begin{array}{cc}
-\gamma(1+r\rho_{0}^{2}\sin2\theta_{0})&-\gamma
r\rho_{0}^{2}(3-\cos2\theta_{0})\\
\gamma
r\rho_{0}^{2}(3+\cos2\theta_{0})&-\gamma(1-r\rho_{0}^{2}\sin2\theta_{0})\\
\end{array}\right).
\end{equation}
As discussed in \cite{Hand}, the type of the stability is
determined by three characteristics of the matrix $(R)$:
\begin{equation}
T=R_{11}+R_{22},~~~d=R_{11}R_{22}-R_{12}R_{21},~~~T^{2}-4d.
\end{equation}
With the help of the matrix $(R)$, it is easy to check, that in
our case the characteristics of the matrix $\|R\|$ are
\begin{equation}
T=-2\gamma
<0,~~~d=\gamma^{2}(1+8r^{2}\rho_{0}^{4})>0,~~~T^{2}-4d=32\gamma^{2}r^{2}\rho_{0}^{2}>0
\end{equation}
and point towards the condition of a stable focus. Note, that the
conditions (39) do not depend on the field parameter
$s=\frac{F}{4\gamma\Omega}$ , and therefore hold for  arbitrary
values of the amplitudes $(A_{1},~A_{2})$.

Thus, in the stationary resonance regime, when frequency of the
external driving field is in the resonance with one of the modes
of the two strongly coupled resonators, the stationary points are
characterized by a stable focus. Therefore, we can argue that the
dissipation leads to a stabilization of the secular instability
regime.

\section{Energy redistribution between resonators}

In this section we will address the problem of the energy redistribution between the resonators $(A_{1}^{2}/A_{2}^{2})$, which are pumped via the
external fields. Let us suppose that the harmonic force acts only on  the second resonator $F_{2}\equiv F,~~F_{1}=0$. Then, with the help of the
expression (10), for the ratio of the oscillation amplitudes
we obtain the following relation:
\begin{equation}
\Bigl|\frac{A_{1}}{A_{2}}\Bigr|=\frac{D}{\tilde{\omega}_{1}^{2}-\Omega^{2}}.
\end{equation}
We recall that the expressions (10) for the amplitudes $A_{1,2}$
with respect to the mode frequencies $\nu_{1,2}$ have the same
resonances embedded in the denominators. They naturally compensate
each other and therefore do not appear in the ratio $A_{1}/A_{2}$.
Nevertheless, as we see from Eq.~(40), another resonance
$\tilde{\omega}_{1}\approx\Omega$ appears in the denominator of
the expression $A_{1}/A_{2}$.

At first we neglect the influence of the damping and the nonlinearity terms, assuming that the frequency of the harmonic force is tuned with one of the partial frequencies of the resonators. Then for the case
when $F_{2}\equiv F,~~\Omega\simeq\tilde{\omega_{2}}$ and
$F_{1}=0$, we obtain
\begin{equation}
\Bigl|\frac{A_{1}}{A_{2}}\Bigr|=\frac{D}{|\tilde{\omega}_{1}^{2}-\tilde{\omega}_{2}^{2}|}\approx\frac{\sigma}{2}.
\end{equation}
Hence, the relation between the amplitudes $A_{1}$ and $A_{2}$ is linear.  From the second resonator a fraction $\frac{\sigma^{2}}{4}$
of the energy is transferred to the first one.
\begin{figure}[t]
 \centering
  \includegraphics[width=10cm]{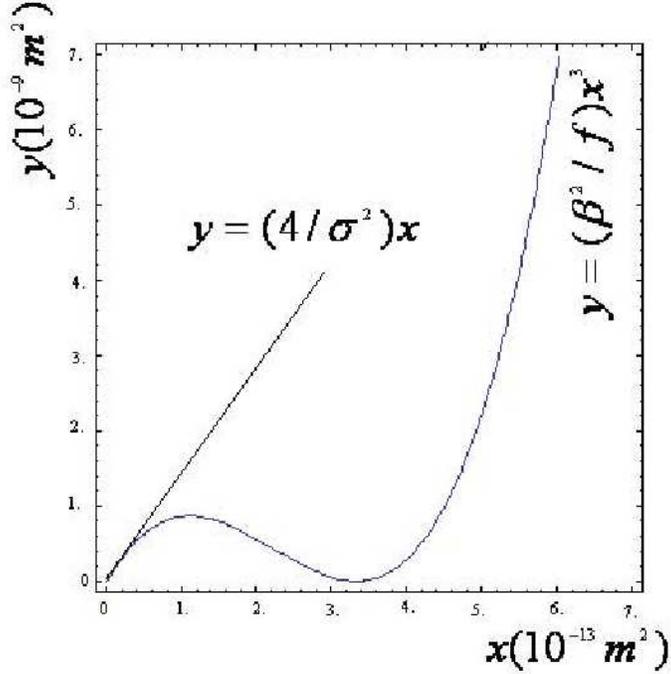}
  \caption{Energy redistribution curve between the coupled resonators and
its asymptotics, plotted using Eq.~(44) for the following values of
parameter $\beta=0.51\cdot10^{20}~Hz/m,~~\Delta
=0.17\cdot10^{8}~Hz,~~\gamma=2.0\cdot10^{5}~Hz,~~f=1.64\cdot10^{10}~Hz^{2}$.}
\end{figure}
The damping and the nonlinear corrections can be considered again with the help of the substitution:
\begin{equation}
\frac{1}{\tilde{\omega}_{1}^{2}-\Omega^{2}}\rightarrow\frac{1}{2\tilde{\omega}_{1}\sqrt{(\beta
A_{1}^{2}-\Delta)^{2}+\gamma^{2}}},~~~\beta=\frac{3}{4}\frac{\chi}{\omega_{1}}.
\end{equation}
From now on we assume that
$$\gamma_{1}\simeq\gamma_{2}\equiv\gamma,~~~\chi_{1}=\chi_{2}=\chi~~~\mbox{and}
~~~\Delta=\omega_{2}-\omega_{1}>0.$$ The resonant denominator is
an important feature of the expression (42). When $A_{1}$ changes
the resonance condition holds in the expression (42). Performing
the substitution (42) in the expression (40) and raising to the
square, we obtain
\begin{equation}
\frac{x}{y}=\frac{f}{(\beta x-\Delta)^{2}+\gamma^{2}} \, ,
\end{equation}
where
$x=A_{1}^{2},~~y=A_{2}^{2},~~f=\frac{D^{2}}{4\tilde{\omega}_{1}^{2}}$.
So, instead of studying the dependence $y=Y(x)$, for convenience
one can convert the problem to one studying the following implicit
function
\begin{equation}
F(x,y)=x[(\beta x-\Delta)^{2}+\gamma^{2}]-fy=0.
\end{equation}
By setting the derivative $dy/dx$  equal to zero, one obtains an equation
for the extrema of the function $y=Y(x)$
\begin{equation}
\frac{dy}{dx}=-\frac{dF/dx}{dF/dy}=f^{-1}(3\beta^{2}x^{2}-4\beta
x\Delta+\Delta^{2}+\gamma^{2})=0.
\end{equation}
It follows then that the points of extremum are
\begin{equation}
x_{1,2}=\frac{2}{3}\frac{\Delta}{\beta}\Bigl(1\pm\sqrt{1-\frac{3}{4}\frac{\Delta^{2}+\gamma^{2}}{\Delta^{2}}}\Bigr).
\end{equation}
For simplicity, we consider the limiting case $\gamma\ll\Delta$.
In this case we get two real roots from Eq.~(46):
\begin{equation}
x_{1}=\frac{\Delta}{\beta}, ~~~x_{2}=\frac{\Delta}{3\beta}.
\end{equation}
It is easy to determine the signs of the second derivatives:
\begin{equation}
\frac{d^{2}y}{dx^{2}}\Big|_{x=x_{1}}=\frac{2\beta\Delta}{f}>0,~~~\frac{d^{2}y}{dx^{2}}\Big|_{x=x_{2}}=-\frac{2\beta\Delta}{f}<0.
\end{equation}
Therefore, the function $y=Y(x)$ has a maximum at the point
$x=x_{2}$, and a minimum at $x=x_{1}$. The curve $y=Y(x)$  is
characterized by two asymptotes as well. The first one, for small
values of $x$ and $y$, is a linear function
$y=(\Delta^{2}/f)x=(4/\sigma^{2})x$. The second one, for large
values of $x$ and $y$, is a cubic function
$y=(\frac{\beta^{2}}{f})x^{3}$. Using the results obtained in this
section, one can plot the curve of the energy redistribution
between the resonators, see Fig.~2.

The anharmonicity  of resonators' oscillations can significantly
change the energy redistribution between the resonators. It turns
out, that the energy pumped into the second resonator via the
external energy source $F_{2}=F$ is transformed to the first
resonator $(F_{1}=0)$ in a different way  depending on the
oscillations amplitude. For small amplitudes of the normal modes,
the energy transfer between the resonators is linear,
$A_{1}^{2}=\frac{\sigma}{4}A_{2}^{2}$ and the transfer rate is
defined by the values of the connectivity $\sigma$.  With
increasing oscillation amplitude the linear law is changed and
turns  nonlinear
$A_{1}^{2}=\Bigl(\frac{4}{9}\cdot\frac{D^{2}}{\chi^{2}}A_{2}^{2}\Bigr)^{1/3}$.
Therefore, we can conclude that the anharmonicity of the
oscillations degrades  the energy transfer rate.

\section{Application to mass measurement sensors and the nonlinear shift of the mode frequencies}

Nanomechanical resonators can be used as apprehensible sensors in
many applications. For a review see \cite{Brueckner}. A decisive
advantage of the nanomechanical resonators are their resonance
frequencies $\omega\approx1GHz$ and quality factors
$Q\approx10^{3}-10^{5}$, which are significantly higher than those
of electrical resonant circuits. That is the reason why
nanomechanical resonators are sensitive transducers for the
detection of molecular systems, in particular for  biological
molecules \cite{Fischer}.  Resonant mass sensor devices operate by
measuring the frequency shift which is proportional to the mass of
the molecules of the material under investigation \cite{Ilic}.
Details of the measurement protocol can be found in \cite{Ekinci}.
Here we briefly refer to the main facts. Assuming that the added
mass $\delta M$ is smaller than the effective resonator mass $M$
one can write a linearized expression $\delta
M\approx\frac{\partial M}{\partial\omega}\delta \omega$. The
minimal measurable frequency shift $\delta\omega$ naturally
defines the sensitivity of the sensor. Due to thermal fluctuations
$\delta\omega>0$. For the single, simple damped harmonic
oscillator system the minimal measurable frequency shift reads
\cite{Ekinci}
\be\delta\omega\approx\Big[\frac{k_{B}T}{M\omega^{2}A^{2}}\frac{\omega\Delta
f}{Q}\Big]^{1/2}.\ee Here $\Delta f $ is the measurement
bandwidth, $M$ is the resonator mass, $\omega$ is the frequency of
the oscillation and $T$ is the temperature. As follows from the
analysis of the preceding sections, at low temperatures the
nonlinear effects (that were not considered in \cite{Ekinci}) can
produce a frequency shift larger than the minimum measurable
frequency shift associated with the thermal effects (see Eq. (18))
$\delta_{1,2}>\omega_{1,2}$. We propose to use the system of
coupled nonlinear oscillators to act as an amplifier for the
frequency shifts. We are convinced that in this way far better
mass measurements are possible for experiments described in
\cite{Ekinci}.

%

\section{Conclusion}

We have developed a general analytical treatment of a system of two coupled driven nonlinear nanomechanical resonators, which is valid for an arbitrary coupling strength (connectivity) between them.
We derive general analytical expressions for  the
amplitude--frequency characteristics of the system with a special
emphasis on the energy redistribution and the energy transport
between the resonators. The obtained results are valid for
arbitrary values of the connectivity.  In the limit of a weak
coupling one recovers the previously obtained results
\cite{Karabalin}. In particular we have shown that the
amplitude--frequency characteristic consists of  two tilted peaks,
the frequency separation between which is equal to twice the value
of the resonators coupling constant $2D$. If the frequency of the
external force $\Omega$ is swept the oscillation amplitude shows
two hysteresis loops in the vicinity of the mode frequencies.
These hysteresis loops contain unstable areas, in which a slight
change of the driving frequency is accompanied by an instantaneous
and a significant change of the oscillation amplitude. This is an
interesting phenomenon, since
 it can be utilized to  switch easily between  the energy transport regimes
of the resonators. We found that for small oscillation amplitudes
 the energy transfer between the resonators follows a linear
law $A_{1}^{2}=\frac{\sigma}{4}A_{2}^{2}$ and the transfer rate is
entirely defined by the values of the
 connectivity $\sigma$. With increasing the oscillation amplitude the energy transfer law
turns  nonlinear
$A_{1}^{2}=\Bigl(\frac{4}{9}\cdot\frac{D^{2}}{\chi^{2}}A_{2}^{2}\Bigr)^{1/3}$
and therefore the transport rate becomes slower. Switching
off the energy transfer rate by tuning of the driving field
frequency is a simple protocol from an experimental point of view
and therefore we expect it to be easily observable.

\textbf{Acknowledgments:}  The financial support by the Deutsche
Forschungsgemeinschaft (DFG) through SFB 762, contract BE
2161/5-1, Grant No. KO-2235/3, the HGSFP (grant number GSC 129/1),
the Heidelberg Center for Quantum Dynamics and STCU Grant No. 5053
is gratefully acknowledged.

\begin{center}
\textbf{Appendix A -- details from section IV}
\end{center}

By taking into account damping effects, the amplitudes of the
forced oscillation of  the nonlinear resonators for an arbitrary
value of the connectivity $\sigma$ are:

\begin{eqnarray}
A_{1,2}=\frac{F_{1,2}(\tilde{\omega}_{2,1}^{2}-\Omega^{2})+F_{2,1}\omega_{-}^{2}\sigma}{4\Omega^{2}}
\cdot\Biggl[\frac{\omega_{+}^{2}}{\omega_{-}^{2}}\frac{1}{\sqrt{1+\sigma^{2}}}\Biggl(\frac{1}{\nu_{1}\sqrt{(\nu_{1}+\delta_{1}-\Omega)^{2}+\Gamma_{1}}}-\nonumber
\end{eqnarray}
\begin{eqnarray}
-\frac{1}{\nu_{2}\sqrt{(\nu_{2}+\delta_{2}-\Omega)^{2}+\Gamma_{2}}}\Biggr)-\Biggl(
\frac{1}{\nu_{1}\sqrt{(\nu_{1}+\delta_{1}-\Omega)^{2}+\Gamma_{1}}}+
\frac{1}{\nu_{2}\sqrt{(\nu_{2}+\delta_{2}-\Omega)^{2}+\Gamma_{2}}}\Biggr)\Biggr].\nonumber
\end{eqnarray}

\begin{center}
\textbf{Appendix B -- details from section V}
\end{center}

The explicit form of the set of equations

\begin{eqnarray}
\frac{dA_{1}}{dt}=\frac{1}{4\Omega}\frac{(1+\sqrt{1+\sigma^{2}})}{\sqrt{1+\sigma^{2}}}F_{1}-\frac{1}{2}
\frac{1}{\sqrt{1+\sigma^{2}}}\Bigl[
\gamma_{1}(1+\sqrt{1+\sigma^{2}})A_{1}+\gamma_{2}(-1+\sqrt{1+\sigma^{2}})A_{1}\Bigr]-\nonumber
\end{eqnarray}
\begin{eqnarray}
-\frac{1}{4\Omega}\frac{1}{\sqrt{1+\sigma^{2}}}\Bigl[\frac{3}{4}\chi_{1}(1+\sqrt{1+\sigma^{2}})(A_{1}^{2}+
A_{2}^{2}+2B^{2})A_{2}+\frac{3}{4}\chi_{2}(-1+\sqrt{1+\sigma^{2}})\times\nonumber
\end{eqnarray}
\begin{eqnarray}
\times\Bigl(\frac{\sigma^{2}}{(1+\sqrt{1+\sigma^{2}})^{2}}(A_{1}^{2}+A_{2}^{2})+
\frac{\sigma^{2}}{(1-\sqrt{1+\sigma^{2}})^{2}}2B^{2}\Bigr)A_{2}\Bigr],\nonumber
\end{eqnarray}
\begin{eqnarray}
\frac{dA_{2}}{dt}=-\frac{1}{2} \frac{1}{\sqrt{1+\sigma^{2}}}\Bigl[
\gamma_{1}(1+\sqrt{1+\sigma^{2}})A_{2}+\gamma_{2}(-1+\sqrt{1+\sigma^{2}})A_{2}\Bigr]+\nonumber
\end{eqnarray}
\begin{eqnarray}
+\frac{1}{4\Omega}\frac{1}{\sqrt{1+\sigma^{2}}}\Bigl[\frac{3}{4}\chi_{1}(1+\sqrt{1+\sigma^{2}})(A_{1}^{2}+
A_{2}^{2}+2B^{2})A_{1}+\nonumber
\end{eqnarray}
\begin{eqnarray}
+\frac{3}{4}\chi_{2}(-1+\sqrt{1+\sigma^{2}})\Bigl(\frac{\sigma^{2}}{(1+\sqrt{1+\sigma^{2}})^{2}}(A_{1}^{2}+A_{2}^{2})+
\frac{\sigma^{2}}{(1-\sqrt{1+\sigma^{2}})^{2}}2B^{2}\Bigr)A_{1}\Bigr],\nonumber
\end{eqnarray}
\begin{eqnarray}
\frac{dB}{dt}=-\frac{1}{4}\Bigl(\frac{-1+\sqrt{-1+\sigma^{2}}}{\sqrt{1+\sigma^{2}}}\gamma_{1}+
\frac{1+\sqrt{1+\sigma^{2}}}{\sqrt{1+\sigma^{2}}}\gamma_{2}\Bigr)B,\nonumber
\end{eqnarray}
\begin{eqnarray}
\frac{d\psi}{dt}=\frac{3}{16\nu_{2}}\cdot\Bigl[\frac{\sigma}{\sqrt{1+\sigma^{2}}}\frac{\sigma}{1+\sqrt{1+\sigma^{2}}}(2A_{1}^{2}+
2A_{2}^{2}+B^{2})\chi_{1}+\nonumber
\end{eqnarray}
\begin{eqnarray}
+\frac{(1+\sqrt{1+\sigma^{2}})}{\sigma}\Bigl(\frac{(1+\sqrt{1+\sigma^{2}})^{2}}{\sigma^{2}}B^{2}+
2\frac{(1-\sqrt{1+\sigma^{2}})^{2}}{\sigma^{2}}(A_{1}^{2}+A_{2}^{2})\Bigr)\chi_{2}\Bigr].\nonumber
\end{eqnarray}

\end{document}